\documentclass[a4paper]{article}

\begin{document}

\title{ $N$-LEVEL QUANTUM SYSTEMS \protect\\
AND LEGENDRE FUNCTIONS\footnote{
{\scriptsize Proceedings of III Seminar}
{\it Nonlinear phenomena in complex systems,}
{\scriptsize Minsk: Institute of Physics, 1995, 328--333.}
\newline \copyright\, 1995 by Institute of Physics, Belarus Academy of Sciences.}}
\author{A.S.~Mazurenko${}^1$,
V.A.~Savva\protect\\
${}^1$ {\footnotesize WWW} {\small address: http://www.geocities.com/amazurenko}
\protect\\
{\it Institute of Physics, Belarus Academy of Sciences }\protect\\
{\it  70 Skorina ave., 220072 Minsk, Republic of Belarus} }
\date{}
\maketitle

\begin{abstract}
An excitation dynamics of new quantum systems of $N$
equidistant energy levels in a mono\-chro\-ma\-tic field has been
investigated. To obtain exact analytical solutions of dynamic
equations an ana\-ly\-tical method based on orthogonal functions
of a real argument has been proposed. Using the orthogonal
Legendre functions we ha\-ve found an exact analytical expression
for a po\-pu\-la\-tion probability amplitude of the level $n$.
Various initial conditions for the excitation of $N$-level
quantum systems have been considered.


\end{abstract}

\section{INTRODUCTION}

Theoretical investigations of the dynamics of the multiphoton
vibrational excitation of polyatomic molecules by an IR laser
radiation are based on the analysis of multilevel quantum systems
as molecular models. Many other problems of nonlinear optics,
spectroscopy and laser physics also lead to the consideration of
the dynamics of multilevel systems. Analytical results obtained up
to now for multilevel systems dynamics do not cover all
the interesting and necessary cases.
\par
In this work an analytical method has been presented that offers
new possibility to obtain exact solutions describing the excitation
dynamics of a number of new multilevel quantum systems in
an infrared laser field.
The method uses orthogonal functions to obtain dynamic equations
solutions.
It generalizes an analytical approach \cite{r1} that is based on integral
transform and orthogonal polynomials.
A particular example with the use of the orthogonal Legendre function
is given. Analytical solutions for two different $N$-level quantum
systems with equidistant levels are found.

\section{ANALYTICAL METHOD}

The dynamics of quantum systems in the monochromatic infrared
radiation $\, {\cal E}_{l}\, \cos(\omega_{l} t)\,$ is described by the
equations
\begin{eqnarray}
 -i \; \frac{da_{n}(t)}{dt} &=& f_{n+1}\;
 e^{-i\epsilon _{n+1}t} \; a_{n+1}(t) +
 f_{n} \; e^{i\epsilon _{n}t} \; a_{n-1}(t)\;, \nonumber \\
\\
a_{n}(t=0) &=& \delta _{n,m} \; ,
\qquad \qquad \qquad n = 0, 1, 2, \ldots  \nonumber
\end{eqnarray}
\noindent for the population probability amplitudes $a_{n}(t)$ of
the level $n$. Here $\;t= \Lambda \tau$ is the dimensionless time
( $\; \Lambda = \mu_{0,1}\;{\cal E}_{l}/(2 \hbar) \;$ is the Rabi
frequency, $\tau$ is ordinary time). Eqs. (1) are obtained from
the Schr\"odinger equation for a multilevel system with the use of
the rotating wave approximation. In such description the
multilevel system has two its characteristics: the dipole moment
$\mu _{n-1,n}$ and the dimensionless frequency detuning
$\varepsilon _{n}$ for every transition $\; n-1 \leftrightarrow
n\;$. The dipole moment function $f_{n}$ describes the dependence
of the radiative transitions moments
\begin{equation}
 \mu _{n-1,n} = \mu _{0,1} \; f_{n}
\end{equation}
\noindent
on the transition number (on energy), $\mu_{0,1}$ is the dipole
moment of the lowest transition. The detuning
\begin{equation}
 \varepsilon _{n} =
 (\omega _{n,n-1} - \omega _{l}) / \Lambda
\end{equation}
\noindent
characterizes the difference of the frequency of the transition $n$
and the laser radiation frequency.
\par
Eqs. (1) for the $N$-level quantum system can be solved by
various methods. In work \cite{r1} analytical method has been
developed on the basis orthonormal polynomial sequences $\;
p_{n}(x)/d_{n}\;$ ($d_{n}$ is a norm). The polynomials satisfy
the recurrence formula
\begin{equation}
 f_{n+1} \; \frac{p_{n+1}(x)}{d_{n+1}} +
 f_{n} \; \frac{p_{n-1}(x)}{d_{n-1}} =
 (rx + s_{n}) \; \frac{p_{n}(x)}{d_{n}} \;.
\end{equation}
\noindent
It is supposed that the coefficients $\; f_{n}\;$ in formula (4)
and in eq. (1) are identical functions, the coefficients $\; s_{n}\;$
and the frequency detuning are connected in the following way
\begin{equation}
 \varepsilon _{n} = s_{n} - s_{n-1} \;.
\end{equation}
\noindent
A number of exact analytical solutions with the help of
the expression
\begin{equation}
 a_{n}(t) = \int \limits _{A}^{B} \sigma (x)\;
 \frac{p_{m}(x)}{d_{m}} \; \frac{p_{n}(x)}{d_{n}}
 \; exp[it(rx + s_{n})] \; dx
\end{equation}
\noindent
has been obtained.
\par
Thus, the po\-ly\-nomials $p_{n}(x)\;$ are orthogonal on the
interval $(A,B)$ with re\-s\-pect to the weight fun\-c\-tion
$\sigma (x)\;$ and give rise to multilevel quantum systems. The
coefficients $f_{n}\;,\; s_{n}\;,\; r\;$ of recurrence formula
(4) define all characteristics of this quantum systems, i.e. the
dipole moment function $f_{n}$ (2), the frequency detuning
$\varepsilon _{n}\;$ (5) and spacing of the energy levels
\begin{equation}
\mbox{{\bf E}}_{n} = \mbox{{\bf E}}_{0} + n\hbar\omega_{l}
 + \hbar\Lambda (s_{n} - s_{0}) \; .
\end{equation}

\section{GENERALIZATION OF ANALYTICAL METHOD}

Let $\varphi_{\alpha} (z)$ are some orthogonal functions of one
variable $z$ (in special case, of course, they can be orthogonal
polynomials). In contrast to polynomials $\; p_{n}(x)\;$ which are
orthogonal with the Kronecker delta $\delta_{m,n} \;$ , the
functions $\varphi_{\alpha} (z)$ can be orthogonal with both the
Kronecker delta and the Dirac $\delta$-function. The scalar
product
\begin{equation}
 D_{\alpha, \beta} \equiv \int \limits _{L}
 \varphi_{\alpha} (z) \; \varphi^{\ast}_{\beta} (z) \;
 dM(z) = \; 0 \; , \;\;\;\;\; \alpha \neq \beta
\end{equation}
\noindent can have infinite values. Here $L$ is an one-dimensional
region of the integration, the measure $dM(z)$ of the integral is
non-negative. $\alpha$ and $\beta$ are real numbers, not only
integer ones. The norm $d_{\alpha} = \sqrt{D_{\alpha, \beta =
\alpha}} \;$ of the orthogonal function can be infinite.
\par
In order to obtain the dynamic equation solution we use
orthogonal functions $\varphi_{n} (z)$ which satisfy
the recurrence formula
\begin{equation}
 f_{n+1} \; \frac{\varphi_{n+1}(z)}{d_{n+1}} +
 f_{n} \; \frac{\varphi_{n-1}(z)}{d_{n-1}} =
 (r \; \theta (z) + s_{n}) \;
 \frac{\varphi_{n}(z)}{d_{n}} \; .
\end{equation}
\noindent
If $\; \theta (z) = x\;$ and $\varphi_{n} (z)$ are the polynomials
$p_{n}(x)$ then (9) becomes (4).
\par
We shall show that any relation ($r, \; A_{\alpha},
\; B_{\alpha}, \; C_{\alpha}$ are constants)
\begin{equation}
 r \; \theta (z) \; \varphi_{\alpha} (z) =
 A_{\alpha} \; \varphi_{\alpha + 1} (z) +
 B_{\alpha} \; \varphi_{\alpha} (z) +
 C_{\alpha} \; \varphi_{\alpha - 1} (z)
\end{equation}
\noindent
for orthogonal functions $\varphi_{\alpha} (z)$ can be reduced to
form (9), when there exists $\; \alpha = \alpha_{0}$ for which
\begin{equation}
 C_{\alpha_{0}} = 0 \; .
\end{equation}
\noindent
Then (10) can be written as
\begin{eqnarray}
 A_{\alpha_{0}} \; \varphi_{\alpha_{0} + 1} (z) &=&
 \left[ r \; \theta (z) - B_{\alpha_{0}} \right] \;
 \varphi_{\alpha_{0}} (z) \; ,                  \nonumber \\
\\
 A_{\alpha_{0} + n} \; \varphi_{\alpha_{0} + n + 1} (z) &=&
 \left[ r \; \theta (z) - B_{\alpha_{0} + n} \right] \;
 \varphi_{\alpha_{0} + n} (z) +
 C_{\alpha_{0} + n} \; \varphi_{\alpha_{0} + n - 1} (z) \; . \nonumber
\end{eqnarray}
\noindent
Without any limitation we can take $\; \alpha_{0} = 0\;$.
Now it is obviously that the function $\varphi_{n} (z)$ has the form
\begin{equation}
 \varphi_{n} (z) = \varphi_{0} (z) \; p_{n}[\theta (z)] \;
\end{equation}
\noindent
where $p_{n}[\theta (z)]$ is a polynomial of the argument $\theta (z)$.
\par
Any orthonormal polynomial $p_{n}(x)/d_{n}$ satisfies the recurrence
formula (4) \cite{r2}. The functions
\begin{equation}
 \varphi_{n} (z) = \varphi_{0} (z) \;
 p_{n}[\theta (z)] / d_{n} \;
\end{equation}
\noindent
also satisfy formula (4) when $\;x = \theta (z)\;$.
And we obtain (9).
Systems of functions with infinite norm $d_{n}$ have to be
considered specially.
\par
Thus, the solution of problem (1) is found in the following
way:
\par
\medskip
\medskip
\medskip
\noindent
\begin{tabular}{p{2.5em}p{34em}}
{\bf (a)} &
  with the help of orthogonal functions
  $\varphi_{\alpha - \alpha_{0}}(z)$ (8) we choose such a sequence
  of functions $\varphi_{n}(z)\;$,
  $\; \alpha - \alpha_{0} \equiv n = 0, 1, 2, \ldots \;$ ,
  which form the complete basis of solutions of the system of
  the relations (12);
  \\[4ex]
{\bf (b)} &
  the recurrence formulae (12) reduce to pattern (9);
  \\[4ex]
{\bf (c)} &
  the coefficients
  \\
\end{tabular}
  \begin{equation}
   \qquad \quad a_{n}(t) = \int \limits _{L} \;
   \frac{\varphi _{m} (z)}{d_{m}} \;
   \frac{\varphi^{\ast}_{n} (z)}{d_{n}} \;
   e^{it(r \theta^{\ast}(z) + s_{n})} \; dM(z) \;
  \end{equation}
\begin{tabular}{p{2.5em}p{34em}}
~ &
  of the expansion of the integral transform kernel \\
\end{tabular}
  \begin{equation}
   \qquad \quad U(t,z) = \frac{\varphi_{m}(z)}{d_{m}} \;
   e^{it(r \theta^{\ast}(z) + s_{n})}
  \end{equation}
\begin{tabular}{p{2.5em}p{34em}}
~ &
  to the series of the functions $\varphi_{n}(z)$ are the solutions
  of problem (1).
  \\[4ex]
\end{tabular}
\par
\noindent
If the expansion of kernel (16) to series of orthogonal
functions (polynomials) $\varphi_{n} (z)$ is known,
the calculation of integral (15) can be avoided.

\section{THE ORTHOGONAL LEGENDRE FUNCTIONS}

Let's give a particular example of orthogonal functions which
satisfy recurrence formula (9). It's known \cite{r3} that
there are systems of functions $\; \varphi^{(N)}_{m}(z)\;$ among
the complete family of the 1st kind Legendre functions $\;
P^{\mu}_{\nu}(z)\;$
\begin{equation}
 \left.
 \begin{array}{l}
  \varphi^{(N)}_{m} (z) = (- 1 )^{N - m -1}
  \Gamma (N - \mu) P^{\mu}_{N - 1} (z)
 \end{array}
 \right| _{\mu = m}
\end{equation}
\noindent
orthonormalized on $[-1,1]$ with the Kronecker $\delta$-function
\begin{equation}
 \int \limits ^{1}_{-1} \;
 \varphi ^{(N)}_{m} (z) \;
 \varphi ^{(N)}_{l} (z) \;
 \frac{dz}{1 - z^{2}}
 = \delta _{m,l} \; d^{2}_{m} \; ,
\end{equation}
\noindent
$m$ and $l$ are non-negative integer number.
$N \geq 1$ is an integer parameter (if $m \geq N$, the right part
limit (17) is calculated at $\mu \rightarrow m$).
The norm of these functions
\begin{equation}
 d_{m} = \left\{
 (N + m - 1)! \; (N - m - 1)! \; /m
 \right\} ^{1/2}
\end{equation}
\noindent
is infinite when $\;m=0\;$.
Therefore the reducing known recurrence formulae \cite{r4} for the
functions $\; P^{\mu}_{\nu}(z)\;$ to relation (9) requires
the use of the linear dependence properties of the Legendre
functions with the integer parameters $\; \mu\equiv m\;$ and
$\; \nu\equiv n\;$.
As a result, the functions (17) satisfy the relation
\begin{equation}
 f_{m+1} \; \frac{\varphi^{(N)}_{m + 1}(z)}{d_{m + 1}} \; +
 \; f_{m} \; \frac{\varphi^{(N)}_{m - 1}(z)}{d_{m - 1}} \; =
 \; r \; \theta (z) \;\; \frac{\varphi_{m}(z)}{d_{m}} \; .
\end{equation}
\noindent
Here $d_{m}$ is norm (18), $\;\theta (z) = z / \sqrt{z^{2} - 1}
\;$, and
\begin{eqnarray}
 f_{m} = r \; \left\{
  \frac{(N - m) \; (N + m - 1)}
                 {m \; (m - 1)}
 \right\} ^{1/2}, \;\;\;\; m>0\;, \nonumber \\
\\
 r = 2 \left\{ (N - 2)(N + 1)/2 \right\} ^{- 1/2}. \nonumber
\end{eqnarray}
\noindent
As in (1), the coefficient $f_{m}$ at $\; m=0\;$ is equal zero.
Since $\; f_{m = N} = 0\;$, we have the system of $N$ linearly
independent recurrence formulae of form (9)
(at $\; s_{n} \equiv 0\;$).
In this work the relation (20) for values $\; m\geq N\;$ is not
considered.

\section{$N$-LEVEL QUANTUM SYSTEMS}

Let's use functions
\begin{equation}
 \varphi ^{(N)}_{m}(z)\;, \;\;\;\;\;
 m \geq 0\;, \;\;\;\;\;
 N \geq 1 \; ,
\end{equation}
\noindent
which form the complete system of solutions of $N$ equations (20)
(when $\; m<N\;$), in order to obtain new solutions of
problem (1).
\par
Let's integrate expression (15) written for functions
$\; \varphi ^{(N)}_{m}(z)\;$ ( $\; m<N\;$ ) which are orthogonal
on $\; [-1,1]\;$ if $\; \theta (z) = z/\sqrt{z^{2}-1}\;$.
\medskip
$$
 a_{n}(t,N) = \int \limits ^{1}_{-1} \;
 \frac{\varphi ^{(N)}_{m} (z)}{d_{n}} \;
 \frac{\varphi^{(N)}_{n} (z)}{d_{n}} \;
  e^{irt(z/\sqrt{z^{2}-1})} \;
 \frac{dz}{1-z^{2}} \; =
$$
\medskip
$$
 =\; 2^{m-n+1}\;\sqrt{\pi}\;
 \left\{
  m\; (N+m-1)!\; (N-m-1)!\;
 \right\} ^{\frac{1}{2}} \;\times
$$
\medskip
$$
 \left\{
 \frac{n\; (N+n-1)!}{(N-n-1)!}
 \right\} ^{\frac{1}{2}}\; e^{irt}\;
 \frac{(irt)^{n+m}}{n!} \; \times
$$
\medskip
\medskip
$$
 \sum_{k=0}^{N-m-1}
 \frac{\Gamma (N-k-\frac{1}{2})\; (-2irt)^{N-m-1-k} }
   {(N+m-k-1)!\; (N-m-k-1)!\; k!\; \Gamma (N+n-k) }
 \; \times
$$
\medskip
\medskip
\begin{equation}
 ~_{2}F_{2}
 \left(
  N+n \; , \;
  n+\frac{1}{2} \; ; \;
  N+n-k \; , \;
  2n+1  \; ; \;
  -2irt
 \right) \;.
\end{equation}
\par
\medskip
\medskip
\par
\noindent
This solution describes the resonant excitation of a
$N$-level quantum system with the equidistant energy spectrum
\begin{equation}
\mbox{{\bf E}}_{n} = \mbox{{\bf E}}_{0} + n\hbar\omega_{l}
\end{equation}
\noindent
where {\bf E}$_{0}\;$ is the zero level energy.
The dipole moment of this system depends on $n$ according to (21).
At the initial time moment $\; t=0\;$ only $m$-level is excited.
\par
Let's obtain another solution of problem (1) for another
dependence on $n$ of the dipole moment function. We can always
consider reverse numeration of levels for finite level quantum
systems, because the choice of "upper" and "lower" level is
arbitrary in such a case. Let's renumber the levels of the quantum
system: $\; n \rightarrow (N-1-n)\;$.
\par
As a result, we obtain an analytical solution with the help of
formulae (21) and (23)
\medskip
\medskip
$$
 a_{n}(t,N) = 2^{m+n-N+2}\;\sqrt{\pi}\;
 \left\{
  m\; (N+m-1)!\; (N-m-1)!
 \right\} ^{\frac{1}{2}} \;\times
$$
\medskip
$$
 \left\{
 \frac{ (N-n-1)\; (2N-n-2)!}{n!}
 \right\} ^{\frac{1}{2}}\; e^{irt}\;
 \frac{(irt)^{N-n+m-1}}{(N-n-1)!}\; \times
$$
\medskip
\medskip
\medskip
$$
 \sum_{k=0}^{N-m-1}
 \frac{\Gamma (N-k-\frac{1}{2})\; (-2irt)^{N-m-1-k} }
   {(N+m-k-1)!\; (N-m-k-1)!\; k!\; \Gamma (2N-n-k-1) }
 \; \times
$$
\medskip
\medskip
\medskip
\begin{equation}
  ~_{2}F_{2}
 \left(
  2N-n-1 ,
  N-n-\frac{1}{2}  ;
  2N-n-k-1 ,
  2N-2n-1  ;
  -2irt
 \right)
\end{equation}
\par
\medskip
\medskip
\noindent
which describes the resonant excitation of a $N$-level quantum
system also with the equidistant energy spectrum, but with
different dipole moment function
\begin{eqnarray}
 f_{n} &=& r \; \left\{
 \frac{n\; (2N - n+1)}
      {(N-n+1) \; (N-n)}
 \right\} ^{1/2}, \;\;\;\; n<N\;, \nonumber \\
\\
 r &=& \left\{ (N - 1)/2 \right\} ^{-1/2}. \nonumber
\end{eqnarray}
\par
Thus, new analytical method for obtaining exact solutions of
the problem of the radiative excitation of multilevel quantum
systems has been proposed. It allows to model the excitation
of systems with more complex dynamics for the description of
which both orthogonal polynomials and orthogonal functions have to
be used.
With the help of the orthogonal Legendre functions exact solutions
for two various $N$-level quantum systems have been obtained.
\bigskip
\bigskip
\par
Financial support from the Belarusian Republican Foundation for
Fundamental Research is acknowledged.
\par

\end{document}